\documentclass[twocolumn]{aastex6}

\fullcollaborationName{Co-authors and affiliations can be found after the acknowledgements}

\begin{document}


\shorttitle{Mg-Poor field stars with SG GC Origins}
\shortauthors{J. G. Fern\'andez-Trincado et al.}

\title{ATYPICAL MG-POOR MILKY WAY FIELD STARS WITH GLOBULAR CLUSTER SECOND-GENERATION LIKE CHEMICAL PATTERNS}

\author{
	J. G. Fern\'andez-Trincado\altaffilmark{1,4},
	O. Zamora\altaffilmark{2,3},
	D. A. Garc\'ia-Hern\'andez\altaffilmark{2,3},	
	Diogo Souto\altaffilmark{5},
	F. Dell'Agli\altaffilmark{2,3},
	R. P. Schiavon\altaffilmark{6}, 
	D. Geisler\altaffilmark{1},
	B. Tang\altaffilmark{1},
	S. Villanova\altaffilmark{1},
	Sten Hasselquist\altaffilmark{7},
	R. E. Mennickent\altaffilmark{1},
	Katia Cunha\altaffilmark{8,5},
	M. Shetrone\altaffilmark{9},
	Carlos Allende Prieto\altaffilmark{2,3},
	K. Vieira\altaffilmark{10},
	G. Zasowski\altaffilmark{11, 12},
	J. Sobeck\altaffilmark{13},
	C. R. Hayes\altaffilmark{13},
	S. R. Majewski\altaffilmark{13},
	V. M. Placco\altaffilmark{14},
	T. C. Beers\altaffilmark{14},
	D. R. G. Schleicher\altaffilmark{1},
	A. C. Robin\altaffilmark{4},
	Sz. M\'esz\'aros\altaffilmark{15, 16},
	T. Masseron\altaffilmark{2,3},
	Ana E. Garc\'ia P\'erez\altaffilmark{2,3},
	F. Anders\altaffilmark{17, 18},
	A. Meza\altaffilmark{19},
	A. Alves-Brito\altaffilmark{20},
	R. Carrera\altaffilmark{2,3},
	D. Minniti\altaffilmark{21,22,23},
	R. R. Lane\altaffilmark{23},
	E. Fern\'andez-Alvar\altaffilmark{24},
	E. Moreno\altaffilmark{24},
	B. Pichardo\altaffilmark{24},
	A. P\'erez-Villegas\altaffilmark{25},
	M. Schultheis\altaffilmark{26},
	A. Roman-Lopes\altaffilmark{27},
	C. E. Fuentes\altaffilmark{1},
	C. Nitschelm\altaffilmark{28},
	P. Harding\altaffilmark{29}, 
	D. Bizyaev\altaffilmark{30, 31},
	K. Pan\altaffilmark{32},
	D. Oravetz\altaffilmark{32},
	A. Simmons\altaffilmark{30},
	Inese I. Ivans\altaffilmark{32},
	S. Blanco-Cuaresma\altaffilmark{33},
	J. Hern\'andez\altaffilmark{34},
	J. Alonso-Garc\'ia\altaffilmark{28, 21},
	O. Valenzuela\altaffilmark{24},
	J. Chanam\'e\altaffilmark{21, 23}\\
	(Affiliations can be found after the references)
}

\altaffiltext{1}{Email: \href{mailto:}{jfernandezt@astro-udec.cl and/or jfernandezt87@gmail.com}}



\begin{abstract}
 We report the peculiar chemical abundance patterns of eleven atypical Milky Way
(MW) field red giant stars observed by the Apache Point Observatory Galactic
Evolution Experiment (APOGEE). These atypical giants exhibit strong Al and N
enhancements accompanied by C and Mg depletions, strikingly similar to those
observed in the so-called second-generation (SG) stars of globular clusters
(GCs). Remarkably, we find low-Mg abundances ([Mg/Fe]$<$0.0) together with
strong Al and N overabundances in the majority (5/7) of the metal-rich
([Fe/H]$\gtrsim - 1.0$) sample stars, which is at odds with actual observations
of SG stars in Galactic CGs of similar metallicities. This chemical pattern is unique 
and unprecedented among MW stars, posing urgent questions about
its origin. These atypical stars could be former SG
stars of dissolved GCs formed with intrinsically lower abundances of Mg and enriched Al (subsequently self-polluted by massive AGB stars) 
or the result of exotic binary systems. We speculate that the stars Mg-deficiency as well as the orbital properties suggest that they could have 
an extragalactic origin. This discovery should guide future dedicated spectroscopic searches of atypical stellar chemical patterns in our Galaxy; 
a fundamental step forward to understand the Galactic formation and evolution.
\end{abstract}



\section{Introduction}
\label{Chapter1}

A number of recent observational studies have revealed that a handful of MW field\footnote{Here the term ``field" refers to stars distributed
across all Galactic components.} stars may exhibit
inhomogeneities in their light-element abundances \citep[e.g.,][]{Carretta2010,
Ramirez2012, Martell2016, Fernandez-Trincado2016b, Schiavon2017a,
Recio-Blanco2017} and neutron-capture element enhancements
\citep[e.g.,][]{Majewski2012, Hasselquist2016, Pereira2017}, similar to those
observed in the SG\footnote{Here we refer to ``SG" as the
groups of stars in GCs that display altered (i.e., different to those of halo
field stars) light-element abundances (He, C, N, O, Na, Al, and Mg).} population
of globular clusters \citep[e.g.,][]{Carretta2009a, Carretta2009b,
Meszaros2015,Carretta2016, Tang2017, Schiavon2017b,
Pancino2017}. 

In this framework, the presence of stars with chemical anomalies in the Galactic field could be explained as the relics of tidally disrupted GCs \citep[e.g.,][and references therein]{Majewski2012, Fernandez-Trincado2016b}, indicating that dissolved GCs could have deposited these eventually unbound stars into the main components of the MW (the bulge, the disk and halo) \citep[e.g.,][]{Carretta2010,  Fernandez-Trincado2013, Kunder2014, Lind2015, Fernandez-Trincado2015a, Fernandez-Trincado2015b, Fernandez-Trincado2016a, Fernandez-Trincado2016b, Martell2016}. 


Despite the enormous progress that has recently been made in exploring abundance anomalies (e.g., C, N, Al) throughout the canonical components of the MW \citep[e.g.,][]{Martell2016, Schiavon2017a}, the distribution and properties of stars originally formed in GCs that are now part of the MW field are still not well understood. Therefore, the study of field stars with ``polluted chemistry" opens a unique window to shed light on models that address the ``mass budget" problem, stellar evolution models, and the phenomenon of multiple populations (MPs) in GCs \citep[see][]{Bastian2015, Ventura2016, Schiavon2017a}. 
Here we report the discovery of atypical MW
field stars with SG-like chemical patterns from the APOGEE survey.

\section{Sample selection}
\label{section2}

Our sample was selected from the APOGEE survey, making use of Sloan Digital Sky Survey-IV (SDSS-IV) Data Release 13 \citep[DR13, ][]{Albareti2016, Majewski2015}. APOGEE DR13 provides chemical and kinematical information of about 150,000 Galactic stars through the analysis of high-resolution ($R\sim$22,500) H-band $\lambda$= 1.51 - 1.69$\mu$m spectra \citep{Zasowski2013}.

We focus our search in the low-metallicity regime ($-1.8<$[Fe/H]$<-0.7$), where stars from the halo and thick disk are expected to dominate the Galactic metallicity distribution \citep[][]{Hawkins2015, Martell2016, Hayes2017}. We impose a mi\-nimal signal-to-noise (S/N) ratio per pixel of 70 to ensure good quality spectra. In order to identify abundance anomalies in MW field stars, we proceed as follows:

From our initial sample (4,611 stars) we selected stars with SG-like chemical patterns in the [Mg/Fe] versus [Al/Fe] plane by means of a clustering analysis. This is done using a k-means clustering approach as described in \citet[][]{Ivezic2014}, with three different centroids in two-dimensional chemical space ([Mg/Fe], [Al/Fe]): i) the SG stars from Galactic GCs ($+0.1, +0.7$); ii) the FG stars in Galactic GCs ($+0.15,-0.2$); and iii) the Galactic thick disk stars ($+0.25, +0.2$). Furthermore, we extended the limits on the Al distribution provided by the k-means analysis for SG-like stars using ge\-nerous Al cuts ([Al/Fe]$\gtrsim +0.1$), and searched for SG-like stars, omitting the carbon-rich stars [C/Fe] $\gtrsim +0.15$ \citep[][]{Schiavon2017a}, which exhibit anomalous chemical abundance patterns as observed in SG GC stellar populations. All the raw data used in this Letter are available in a public repository\footnote{\url{https://github.com/Fernandez-Trincado/ChemicalAnomalies/blob/master/README.md}}.

Figure \ref{Figure1}a shows the locus occupied by our SG-like candidates,  which are located above the dashed grey line that was derived according to the {k-means} algorithm. Stars from Galactic GCs of similar metallicity \citep{Meszaros2015} and the N-rich field stars of \citet{Martell2016, Schiavon2017a} are also indicated in the figure for illustration.  Indeed, ten of the {N-rich} stars reported by \citet{Schiavon2017a} are situated in the locus of SG-like stars found by the k-means algorithm.

After applying the criteria cited above, we have a sub-sample of 260 stars, from which 58.5\% (152/260) are known stars from clusters and other anomalous stars previously reported in the literature \citep{Meszaros2015, Fernandez-Trincado2016a, Tang2017, Schiavon2017a}, and 28.5\% (74/260) have no significant N overabundances (see \S\ref{section3}) and were rejected.

To discard false positives in the remaining 34 stars, the most relevant atomic (Al, Mg, Si, and Ni) and molecular (CO, CN, and OH) spectral features in the H-band were visually inspected, 
to ensure that the final APOGEE spectra are of good quality (e.g., not critically affected by detector persistence, proper continuum normalization, telluric- and sky- lines correction, etc.), 
to provide reliable chemical abundances. We end with a final sample comprising eleven stars (Table \ref{Table1}).


\section{Chemical abundance analysis}
\label{section3}

\begin{figure*}
	\begin{center}		
		\includegraphics[width=0.7\textwidth]{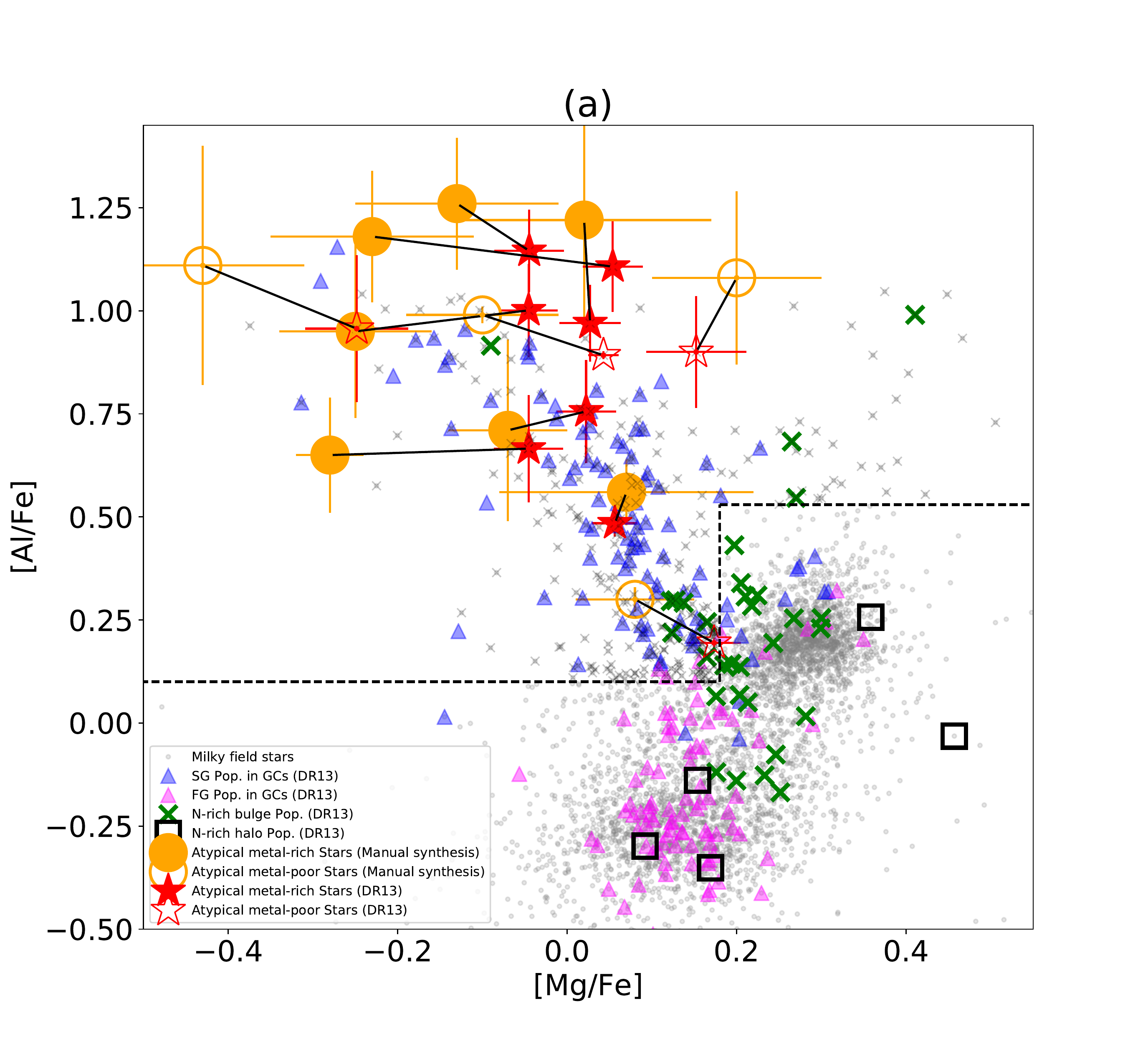}
		\includegraphics[width=0.52\textwidth]{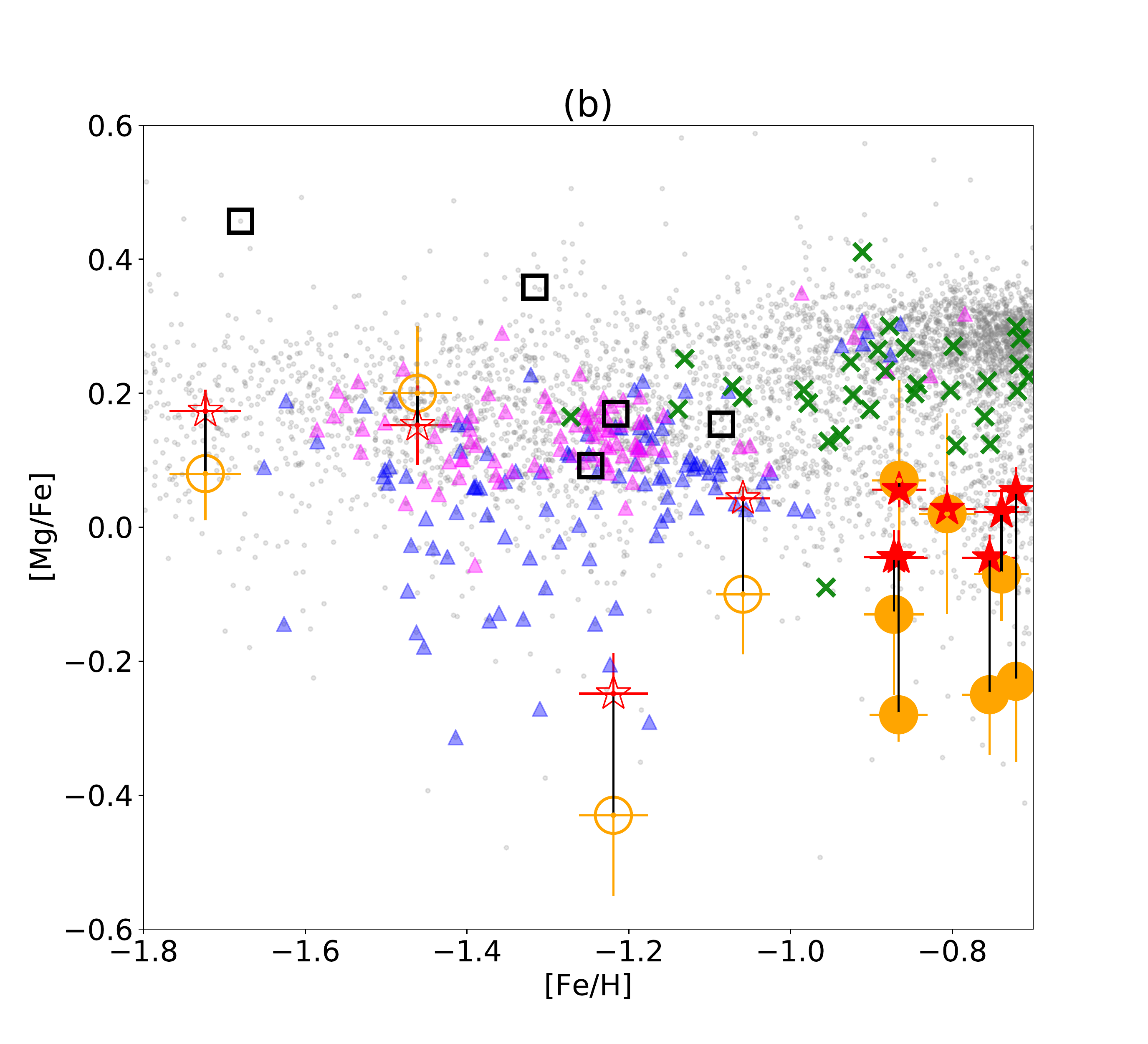}\includegraphics[width=0.52\textwidth]{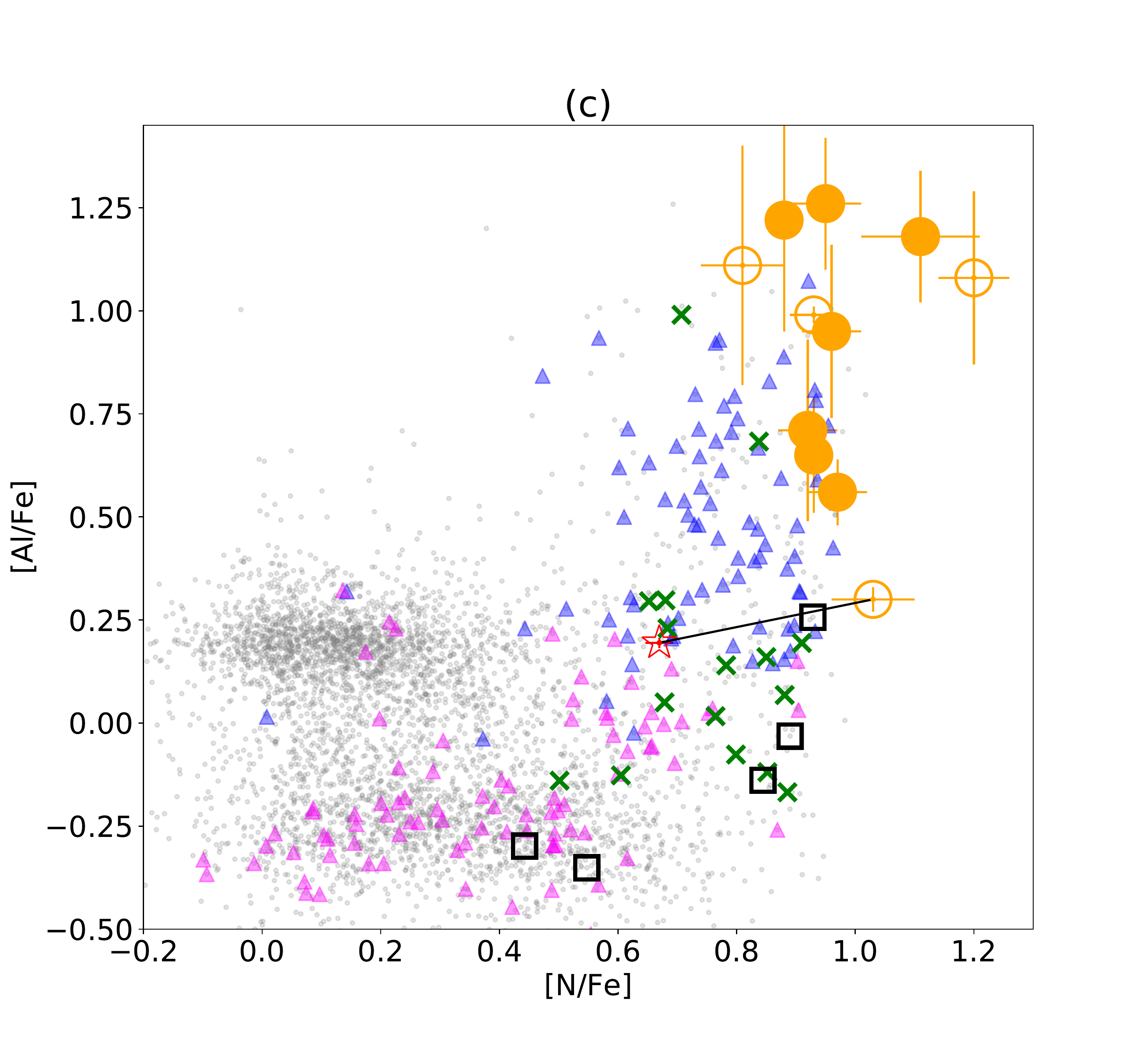}
	\end{center}
	\caption{Abundance ratios in three different planes: (a) [Mg/Fe]-[Al/Fe],
(b) [Fe/H]-[Mg/Fe], and (c) [N/Fe]-[Al/Fe], for the new field SG GC-like stars
(red star symbols for DR13 abundances, orange circles for our manual analysis) overlaid with MW field stars, N-rich halo stars
\citep[][]{Martell2016}, N-rich bulge stars \citep[][]{Schiavon2017a}, FG and SG populations in GCs M2, M3, M5, M107, M71 and M13 \citep{Meszaros2015}. Open circles indicate the SG-like candidates with
[Fe/H]$< -1$. In (a) the grey dashed line marks the loci of the SG GC-like
candidates, above [Al/Fe]$>+0.1$ ([Mg/Fe]$<+0.18$) and [Al/Fe]$>+0.53$
([Mg/Fe]$> +0.18$ ), based on k-means clustering.} 
	\label{Figure1}
\end{figure*}

We have analyzed up to nine chemical elements that are typical indicators of the presence of SG stars in GCs (C, N, O, Mg, Al, Si, Ni, Na, and Fe). The APOGEE DR13 does not provide reliable N abundances for most of our potential candidates because they show very strong CN lines, falling near the high-N edge of the grid and consequently flagged as ``GRIDEDGE\_BAD" in DR13 (except 2M02491285$+$5534213 with [N/Fe]=$+0.67$, see Figure \ref{Figure1}c).

\begin{figure*}
	\begin{center}
		\includegraphics[width=160mm]{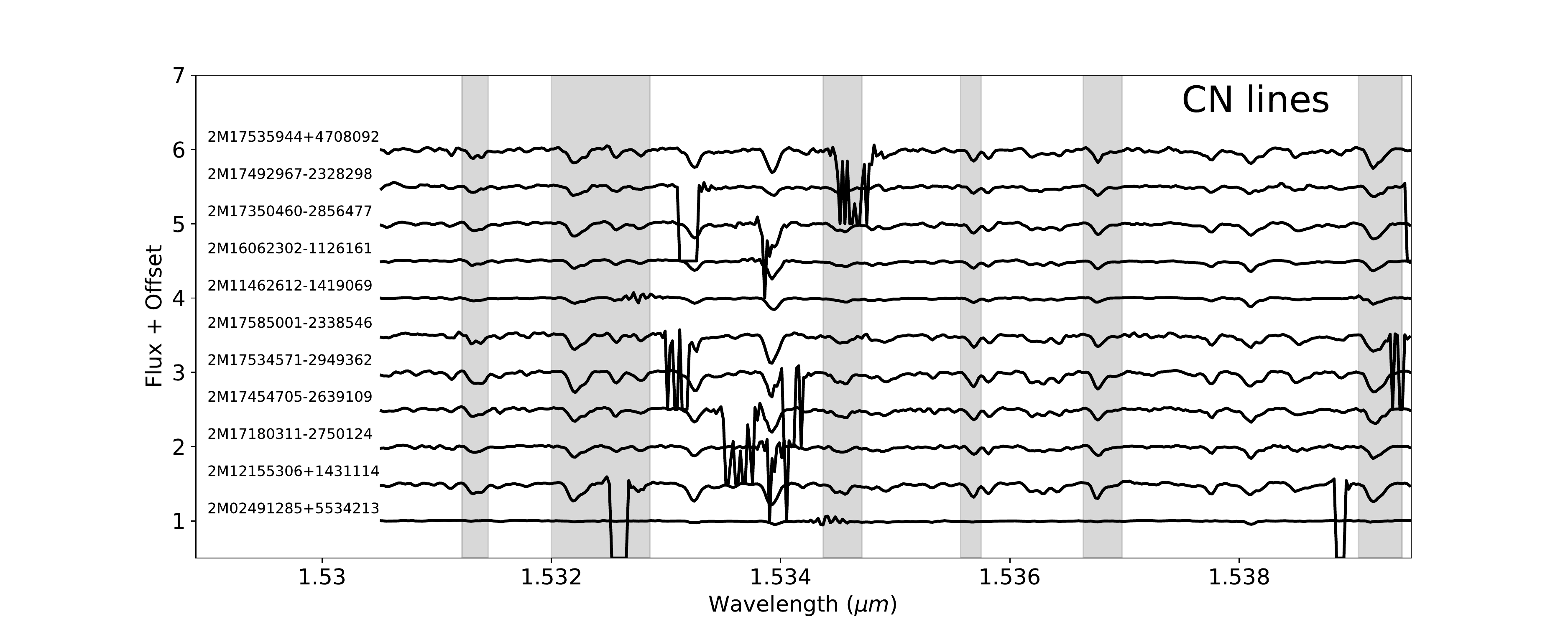}
		\includegraphics[width=160mm]{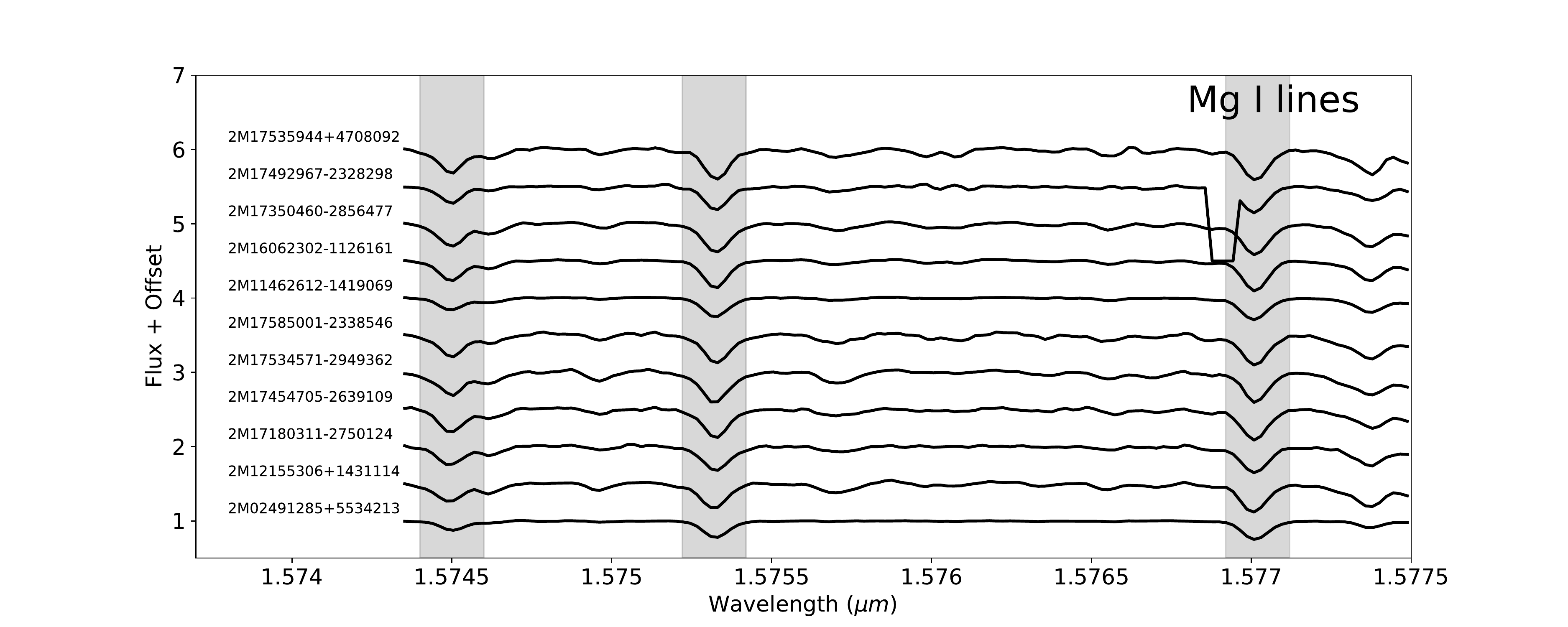}
		\includegraphics[width=160mm]{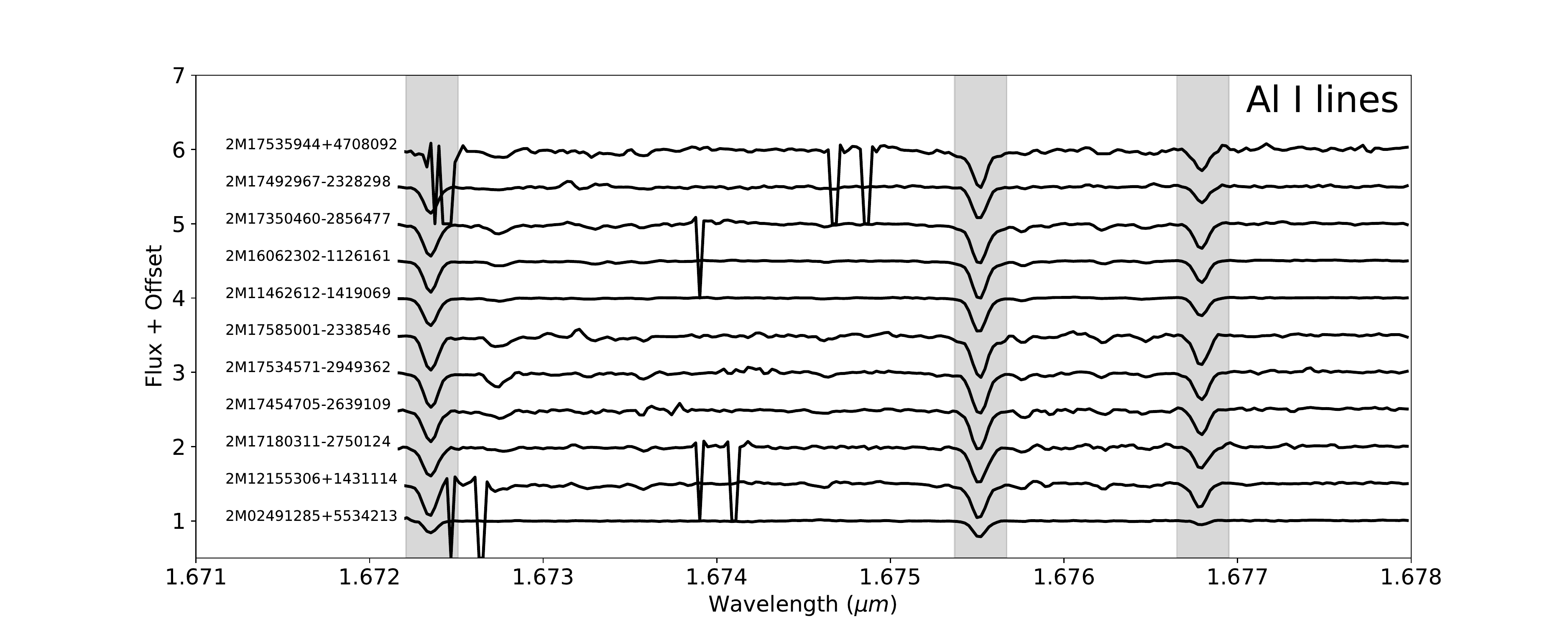}
		\caption{The H-band spectra of our atypical field stars,
covering spectral regions around CN bands, Mg\,{\sc i}, and Al\,{\sc i}. The grey vertical bands indicate some of the wavelenght regimes of the spectral features used in our analysis. The spectra have been shifted to a common wavelength scale.}
		\label{Figure2}
	\end{center}
\end{figure*}

In order to provide a consistent chemical analysis, we re-determine the
chemical abundances by means of a line-by-line analysis. The chemical abundances
have been derived assuming as input the effective temperature (T$_{\rm eff}$)
and metallicity as derived by the APOGEE Stellar Parameter and Chemical
Abundances Pipeline \citep[ASPCAP;][]{GarciaHernandez2016}. However, we do not
adopt the surface gravity (log g) provided by ASPCAP, since it is affected by a
systematic effect that overestimates the log g values (Holtzman et al., in
preparation). We estimate surface gravity from 10 Gyr PARSEC \citep{Bressan2012}
isochrones \citep[10 Gyr is the typical age of Galactic GCs;][]{Harris2010}. The
line list used in this work is the latest internal DR13 atomic/molecular
linelist (linelist.20150714), and the line-by-line analysis was done using the
1D spectral synthesis code Turbospectrum \citep{Alvarez1998} and MARCS model
atmospheres \citep{Gustafsson2008}. In particular, a mix of heavily CN-cycle and
$\alpha$-poor MARCS models were used. The same molecular lines adopted by
\citet{Smith2013} and \citet{Souto2016} were employed to determine the C, N, and
O abundances. Examples for a portion of the observed APOGEE spectra (spectral
region covering CN, Mg, and Al lines) are shown in Figure \ref{Figure2} for our
eleven anomalous stars. Table \ref{Table1} lists the final set of atmospheric
parameters and chemical abundances for each star obtained through ASPCAP DR13
(first line), and the line-by-line synthesis calculations adopting log g from
theoretical isochrones and using the tools mentioned above (second line).

We find the differences in the star-to-star abundances between ASPCAP DR13
and our manual analysis to be small, $\Delta$[Mg/Fe]$\lesssim+0.2$,
$\Delta$[Al/Fe]$\lesssim+0.15$, $\Delta$[O/Fe]$\lesssim+0.2$,
$\Delta$[Si/Fe]$\lesssim+0.15$, and $\Delta$[Ni/Fe]$\lesssim+0.15$, generally
overlapping with our internal errors. It is important to note that these
discrepancies do not affect the main conclusion of this work, i.e., both
line-to-line abundances and DR13 abundances indicate that these stars are N-rich
and Al-rich. Mg abundances are usually lower in the manual analysis compared with ASPCAP, a result already found in similar type of SG-like field stars
\citep{Fernandez-Trincado2016b}. We note that Na abundances are more discrepant
between DR13 and our manual analysis. As the Na lines are usually weak
(especially in the most metal-poor stars; [Fe/H] $<$ $-$1.0), the uncertainty in
the Na abundance is strongly modulated by the uncertainty in the continuum
location. ASPCAP uses a global fit to the continuum in three detector chips
independently, while we place the pseudo-continuum in a region around the lines
of interest. We believe that our manual method is more reliable, since it avoids
possible shifts in the continuum location due to imperfections in the spectral
subtraction along the full spectral range. This way, our manual analysis shows
the Na-rich nature of the SG-like candidates.

\section{Orbital information}
\label{section5}

We use the galactic dynamic software \textit{GravPot16}\footnote{\url{https://fernandez-trincado.github.io/GravPot16/}} \citep[Model 4 in][]{Fernandez-Trincado2016b} to predict the trajectories for five stars (Table \ref{Table2}), from which the space velocity and position vectors can be fully resolved. 


To construct the stellar orbits we employed radial velocities derived from
APOGEE DR13, proper motions from UCAC-5
\citep[][]{Zacharias2017}, and APOGEE distance estimates from
\citet[][]{Santiago2016} and \citet{Anders2017}. The orbital elements are
listed in Table \ref{Table2}.

All five stars indeed lie on very eccentric orbits (\textit{e}$>$0.65) passing through
the Galactic bulge, reflecting a potentially unusual origin in the MW.

In particular, two stars (2M17535944$+$4708092 and 2M12155306$+$1431114)
have relatively high metallicity ([Fe/H]$\sim-0.8$) and may reach distances of
up to $Z_{max}\sim$ 17 kpc above the Galactic plane.

These orbital properties (together with the unusually low levels of Mg observed
in the most metal-rich stars) may support our speculated scenario discussed below, in which these atypical stars may have an extragalactic origin.


\section{Discussion} 
\label{section4}

The main finding of this work is the discovery of eleven atypical MW
field red giant stars with SG GC-like abundance patterns; i.e., with strong
enrichments in N, Na, Si, and Al, accompanied by decreased abundances of C, O, Ni, and Mg. 
Figure \ref{Figure1}b shows
that most of the new chemically anomalous stars exhibit significantly lower
[Mg/Fe] ratios (at [Fe/H]$\gtrsim-1.0$) as compared to Galactic disk
stars (at the same metallicity) and the N-rich halo and bulge stars \citep[e.g.,][]{Martell2016, Schiavon2017a}. This suggests
that the vast majority of our stars have an unusual origin. The exceptions are the two most metal-poor stars ([Fe/H]$\lesssim-1.4$), which display higher [Mg/Fe] ratios similar to the ``canonical halo". Their [Al/Fe] and [N/Fe] ratios, however, are significantly higher than those of the bulk of MW field stars (Figures \ref{Figure1}a and \ref{Figure1}c),
indicating that they may be SG stars originally formed from material that was
chemically enriched in GCs \citep[][]{Martell2016, Schiavon2017a}. For
example, the measured abundances are in nice agreement with the pollution
expected by massive AGB stars at metallicity lower than [Fe/H]$<-1.4$
\citep[][Dell'Agli et al. 2017 in prep.]{Ventura2016}.

Interestingly, the most-metal rich ([Fe/H]$\gtrsim-1.2$) and atypical Mg-poor
stars appear to belong to two groups, according to their Fe abundance (see
Figure \ref{Figure1}). A first group, only two stars with
$-1.2\lesssim$[Fe/H]$\lesssim-1.0$, exhibit Mg depletion more or less consistent
(within the errors) with the Mg abundances typically observed in Galactic GCs of
similar metallicities \citep[][]{Meszaros2015}. The second group (7 stars),
however, displays similar Mg depletion (Figure \ref{Figure1}), but at higher
metallicities ([Fe/H]$>-1.0$). This Mg-deficiency ([Mg/Fe]$\lesssim$0) --
coupled with strong N and Al enrichment ([N,Al/Fe]$\gtrsim+0.5$) -- is at odds
with present observations of SG stars in Galactic GCs of similar metallicities
(Figure \ref{Figure1})\footnote{To our knowledge, NGC 2419 ([Fe/H]$\sim$-2.0) is
the only Galactic GC where many SG stars with very low Mg have been detected
\citep[see e.g.,][]{Ventura2012}. Because of NGC 2419's complex chemistry ,
several authors have indeed suggested that NGC 2419 has an extragalactic origin
\citep[see e.g.,][]{Cohen2010, Cohen2011, Cohen2012,Mucciarelli2012}.}. In addition, Figure
\ref{Figure1} shows that this Mg-deficiency is not seen in the vast majority of
N-rich bulge stars of similar metallicity \citep[][]{Schiavon2017a}; only one
N-rich bulge star displays a chemical pattern identical to the atypical stars
reported here (Figure \ref{Figure1}). A total of six atypical sample stars are seen to lie towards the bulge but is not clear if they could be (or not) some kind related to
the latter N-rich bulge population. 

Could these atypical stars be chemically tagged as migrants from dwarf galaxies?
We find this possibility unlikely because our stars display [Al/Fe] much higher
than observed in dwarf galaxy stellar populations today
\citep[e.g.,][]{Shetrone2003, Hasselquist2017}. However, these stars could be
former members of a dwarf galaxy (with intrinsically lower Mg) polluted by a
massive AGB star in a binary system, which could produce the
chemical pattern observed. Such an
exotic binary system seems to be unlikely. Indeed, no star in our sample
exhibits significant photometric and/or radial velocity variability (see Table
\ref{Table2}). Follow-up observations (e.g., more radial velocity data) would confirm/disprove the
binary hypothesis.

Recently, \citet{Ventura2016} has reported a remarkable agreement between
the APOGEE Mg-Al anticorrelations (two elements sensitive to the metallicity of
the GC polluters) observed in Galactic GCs ($-2.2\lesssim$[Fe/H]$\lesssim-1.0$)
and the theoretical yields from massive AGB stars (m-AGBs). This further
supports the idea that SG-GC stars formed from the winds of m-AGBs, possibly
diluted with pristine gas with the same chemical composition of the FG stars
\citep[see also][]{Renzini2015}. At higher metallicities $-1 <$ [Fe/H] $< -0.7$,
however, the maximum Al spread (with respect to the FG) expected from the
ejecta of m-AGBs is in the range $+0.2 <$ $\Delta$(Al) $<$ $+$0.5
\citep[][Dell'Agli et al. 2017 in prep.]{Ventura2016} but only a modest Mg
depletion is expected. The high Al observed ([Al/Fe]$\gtrsim$$+0.6$) in the
atypical stars at these metallicities could be explained under the m-AGBs
pollution framework if they are earlier SG members of dissolved GCs
\citep[see][]{Schiavon2017a} where the FG stars formed with higher levels of Al.
The FG stars in metal-rich ([Fe/H]$\gtrsim-1.0$) Galactic GCs such as M 107, M
71, 47 Tuc, and NGC 5927 \citep[][]{Meszaros2015, Pancino2017} are known to
be formed with a higher Al (compared to a purely solar-scaled mixture); but both FG and SG stars exhibit similarly high Mg abundances - with no significant
spread between the two stellar generations, as predicted by the m-AGBs
self-enrichment scenario \citep[][Dell'Agli et al. 2017 in prep]{Ventura2016}.

Therefore, the chemical composition of our atypical metal-rich stars,
particularly the observed Al overabundances coupled with low Mg, cannot be
explained by invoking pollution from m-AGBs alone (formed with
a solar-scaled or an $\alpha$-enhanced mixture).
A possible explanation for these chemical
anomalies is that these stars escaped from GCs whose FG stars formed with a
chemical composition enriched in Al but with a lower Mg content in comparison
with the standard solar-scaled or $\alpha$-enhanced mixture. This could be
obtained if we hypothesize that the gas cloud from which the GC formed was
mainly polluted by SN explosions of stars of about $\sim$20$-$30 M$_\odot$,
characterized by medium or large rotation rates during their life, according to
the most recent yields by Limongi \& Chieffi (2017, in prep.). Under these
conditions the gas ejected is expected to be slightly enriched in Al but
Mg-poor. If the FG stars formed with this chemistry, then subsequent pollution
from m-AGBs would form SG stars with the same chemical composition of
the atypical Mg-poor SG-like stars reported here.

[Mg/Fe] (or [Mg/$\alpha$]) from high-resolution integrated-light
spectroscopic observations in extragalactic GCs -- even with average
metallicities similar to our atypical Mg-poor stars -- is generally lower than
in Galactic GCs with similar metallicity \citep[e.g.,][]{Pancino2017}.  A low [Mg/Fe] ratio coupled with high Al (when
available) is also observed in some extragalactic GCs \citep[e.g., in M 31 and
LMC GCs; see e.g.,][]{Colucci2009, Colucci2012}. At present, possible
explanations for the low Mg content in some extragalactic GCs include both
internal and external effects, which could also work simultaneously
\citep[e.g.,][]{Pancino2017}. The internal effect is linked to the particular
formation and chemical evolution of a given GC (e.g., NGC 2419), while the
external effect is related to the specific chemical evolution of their host
galaxies. 

In short, the unique Mg-deficiency of the discovered
atypical metal-rich stars with SG-like chemical patterns
(as well as their orbital properties) suggest that these
stars may have an extragalactic origin; e.g., they could
be former members of dissolved extragalactic GCs, the
remnants of stellar systems accreted long time ago by
our Galaxy. This finding should encourage future dedicated searches (e.g., with on-going massive spectroscopic surveys 
like APOGEE-2, Gaia-ESO, etc.) of chemically atypical Galactic stars, something that would 
represent a major advance to understand the formation and evolution of our own Galaxy.

\clearpage
\floattable
\begin{deluxetable*}{ccccccccccccc}
	\center
	\tabletypesize{\small}
	\tablecolumns{12}
	\tablewidth{0pt}
	\rotate
	\tablecaption{Abundance anomalies identified in this study}
	\tablehead{
		\colhead{\bf APOGEE ID} &    \colhead{T$_{\rm eff}$} &    \colhead{log g}   & \colhead{ [Fe/H] } & \colhead{[C/Fe]}  & \colhead{[N/Fe]} & \colhead{[O/Fe]} &  \colhead{[Al/Fe]} & \colhead{[Mg/Fe]} &  \colhead{[Si/Fe]}  & \colhead{[Ni/Fe]} & \colhead{[Na/Fe]}\\
		\colhead{} &   \colhead{(K)} &   \colhead{(dex)}    &  \colhead{(dex)}  & \colhead{(dex)}   & \colhead{(dex)} & \colhead{(dex)} &  \colhead{(dex)} &  \colhead{(dex)} & \colhead{(dex)} & \colhead{(dex)} & \colhead{(dex)}
	}
	\startdata
	2M17535944+4708092  & 4154.6  &  1.36    &  -0.86$\pm$0.03   &  -0.05 $\pm$ 0.02 &  ...              &    0.19 $\pm$  0.02 &  0.48 $\pm$ 0.03 &   0.06 $\pm$ 0.03 &  0.46 $\pm$ 0.02 &  -0.09 $\pm$ 0.01  &  -0.40 $\pm$ 0.20 \\
			&         &  0.96    &                 &  -0.13$\pm$0.03  &  0.97$\pm$0.05  &  0.34$\pm$0.05 &  0.56$\pm$0.08 &   0.07$\pm$0.15 &     0.59$\pm$0.02 &  -0.11$\pm$0.03 &  0.26$\pm$0.08\\    \hline
	2M17585001-2338546  & 4169.4  &  1.63    &  -0.75$\pm$0.03 &  -0.27 $\pm$ 0.05 &  ...              &   -0.02 $\pm$  0.04 &  1.00 $\pm$ 0.11 &  -0.05 $\pm$ 0.03 &  0.16 $\pm$ 0.04 &   0.02 $\pm$ 0.02  &  -0.07 $\pm$ 0.16 \\
		&         &  1.06    &                 &  -0.30$\pm$0.03  &  0.96$\pm$0.05  &  0.17$\pm$0.06 &  0.95$\pm$0.21 &  -0.25$\pm$0.09 &     0.32$\pm$0.05  & 0.07$\pm$0.05 & 0.31$\pm$0.05\\ \hline
	2M17350460-2856477  & 4218.9  &  1.66    &  -0.74$\pm$0.03 &  -0.33 $\pm$ 0.06 &  ...              &    0.02 $\pm$  0.04 &  0.76 $\pm$ 0.13 &   0.02 $\pm$ 0.04 &  0.18 $\pm$ 0.04 &   0.05 $\pm$ 0.02  &   0.04 $\pm$ 0.18 \\
		&         &  1.13    &                 &  -0.20$\pm$0.04  &  0.92$\pm$0.05  &  0.26$\pm$0.04 &  0.71$\pm$0.22 &  -0.07$\pm$0.07 &     0.31$\pm$0.05 & 0.08$\pm$0.06 & 0.36$\pm$0.02\\ \hline
	2M12155306+1431114  & 4279.5  &  1.59    &  -0.87$\pm$0.04 &  -0.04 $\pm$ 0.07 &  ...              &    0.11 $\pm$  0.05 &  0.67 $\pm$ 0.13 &  -0.05 $\pm$ 0.04 &  0.21 $\pm$ 0.04 &   0.04 $\pm$ 0.03  &  -0.11 $\pm$ 0.19 \\
	&         &  1.08    &                 &  -0.22$\pm$0.04  &  0.93$\pm$0.03  &  0.07$\pm$0.04 &  0.65$\pm$0.14 &  -0.28$\pm$0.04 &     0.29$\pm$0.10 & -0.004$\pm$0.06 & 0.229$\pm$0.07\\  \hline
	2M16062302-1126161  & 4325.8  &  1.50    &  -1.06$\pm$0.03 &  -0.41 $\pm$ 0.01 &  ...              &    0.09 $\pm$  0.01 &  0.89 $\pm$ 0.01 &   0.04 $\pm$ 0.02 &  0.26 $\pm$ 0.01 &  -0.01 $\pm$ 0.00  &  ... \\
		&         &  1.07    &                 &  -0.30$\pm$0.07  &  0.93$\pm$0.04  &  0.17$\pm$0.04 &  0.99$\pm$0.02 &  -0.10$\pm$0.09 &     0.38$\pm$0.01 & 0.002$\pm$0.01 &  ... \\ \hline
	2M17454705-2639109  & 4419.2  &  1.85    &  -0.81$\pm$0.04 &  -0.23 $\pm$ 0.06 &  ...              &    0.08 $\pm$  0.05 &  0.97 $\pm$ 0.09 &   0.03 $\pm$ 0.04 &  0.23 $\pm$ 0.02 &   0.04 $\pm$ 0.02  &  -0.58 $\pm$ 0.26 \\
		&         &  1.36    &                 &  -0.11$\pm$0.01  &  0.88$\pm$0.03  &  0.24$\pm$0.08 &  1.22$\pm$0.27 &   0.02$\pm$0.15 &     0.34$\pm$0.04 & 0.03$\pm$0.04 & 0.33$\pm$ ...\\  \hline
	2M17492967-2328298  & 4428.5  &  1.77    &  -1.46$\pm$0.04 &   0.03 $\pm$ 0.09 &  ...              &    0.01 $\pm$  0.07 &  0.90 $\pm$ 0.14 &   0.15 $\pm$ 0.06 &  0.33 $\pm$ 0.05 &   0.11 $\pm$ 0.03  &   ... \\
		&         &  0.91    &                 &  -0.31$\pm$0.03  &  1.20$\pm$0.06  &  0.09$\pm$0.04 &  1.08$\pm$0.21 &   0.20$\pm$0.10 &     0.41$\pm$0.04 & 0.17$\pm$0.04 &  ...\\   \hline
	2M17534571-2949362  & 4484.7  &  2.18    &  -0.72$\pm$0.03 &   0.07 $\pm$ 0.06 &  ...              &    0.13 $\pm$  0.05 &  1.11 $\pm$ 0.11 &   0.05 $\pm$ 0.04 &  0.26 $\pm$ 0.04 &   0.03 $\pm$ 0.02  &   0.11 $\pm$ 0.17 \\
		&         &  1.53    &                 &  -0.12$\pm$0.07  &  1.11$\pm$0.10  &  0.14$\pm$0.03 &  1.18$\pm$0.16 &  -0.23$\pm$0.12 &     0.39$\pm$0.07 & 0.04$\pm$0.04 & 0.43$\pm$0.04\\ \hline
	2M11462612-1419069  & 4564.8  &  1.79    &  -1.22$\pm$0.04 &  -0.39 $\pm$ 0.11 &  ...              &   -0.16 $\pm$  0.10 &  0.96 $\pm$ 0.18 &  -0.25 $\pm$ 0.06 &  0.30 $\pm$ 0.06 &   0.01 $\pm$ 0.05  &   ... \\
		&         &  1.32    &                 &  -0.30$\pm$0.06  &  0.81$\pm$0.07  & -0.03$\pm$0.04 &  1.11$\pm$0.29 &  -0.43$\pm$0.12 &     0.41$\pm$0.07 & -0.06$\pm$0.05 & ... \\ \hline
	2M17180311-2750124  & 4725.3  &  2.19    &  -0.87$\pm$0.04 &  -0.13 $\pm$ 0.07 &  ...              &    0.15 $\pm$  0.08 &  1.15 $\pm$ 0.10 &  -0.04 $\pm$ 0.04 &  0.32 $\pm$ 0.05 &   0.09 $\pm$ 0.03  &   0.29 $\pm$ 0.18 \\
		&         &  1.73    &                 &  -0.09$\pm$0.02  &  0.95$\pm$0.06  &  0.01$\pm$0.03 &  1.26$\pm$0.16 &  -0.13$\pm$0.12 &     0.35$\pm$0.04 & 0.03$\pm$0.05 & 0.65$\pm$ ...\\ \hline
	2M02491285+5534213  & 4762.3  &  2.06    &  -1.72$\pm$0.04 &   0.09 $\pm$ 0.01 &  0.67 $\pm$ 0.02  &    0.22 $\pm$  0.03 &  0.19 $\pm$ 0.01 &   0.17 $\pm$ 0.03 &  0.19 $\pm$ 0.01 &   0.14 $\pm$ 0.02  &   ... \\
		&         &  1.44    &                 &  -0.18$\pm$0.01  &  1.03$\pm$0.07  &  0.12$\pm$0.07 &  0.30$\pm$0.03 &   0.08$\pm$0.07 &     0.31$\pm$0.04 & 0.17$\pm$0.03 & ...\\
	\enddata
	\tablecomments{The first and second rows show the DR13 and our manual
	results, respectively. The Na abundances for the [Fe/H] $<$ $-$1.0 stars
	(not listed) are not not reliable.}
		\label{Table1}
\end{deluxetable*}

\floattable
\begin{table*}
	\setlength{\tabcolsep}{2.0mm}  
	\caption{Variations between 2MASS and DENIS magnitudes and radial velocities ($\sigma RV$) over the period of the APOGEE observations. Columns 5, 6, 7, and 8 show the median perigalactic distance, the median apogalactic distance, the median maxium distance from the Galactic plane, and the median eccentricity, respectively.}
	\label{Table2}
	\begin{tabular}{cccccccc}
		\hline
		\hline
		{\bf APOGEE ID} &   ${ K_{2MASS}} - { K_{DENIS}}$ & $N_{visits}$ & $\sigma RV$ & median $r_{peri}$ &  median $r_{apo}$ &  median $Z_{max}$ & median $e$ \\  
		                         &         (mag)        &  & km s$^{-1}$ & kpc & kpc & kpc & \\  
		\hline
		\hline
		2M17535944+4708092   &  ...             &   3  &      0.21    & 3.84$^{+4.3}_{-2.4}$ &  21.54$^{+48}_{-5.5}$ & 18.79$^{+43.4}_{-7.6}$ & 0.76$^{+0.16}_{-0.19}$ \\
		2M17585001-2338546   &  -0.064      &  1   &      ...        & ... & ... & ... & ... \\
        2M17350460-2856477   &  0.134        &  2   &      0.23    & ... & ...  & ...  & ... \\		
        2M12155306+1431114   &  ...               &  13  &      0.13  & 4.078$^{+3.8}_{-2.9}$ & 17.6$^{+29}_{-1.9}$ & 16.04$^{+17.41}_{-1.9}$ & 0.69$^{+0.2}_{-0.15}$ \\
		2M16062302-1126161   &  -0.071       &  4   &      0.24   & 1.15$^{+0.49}_{-0.76}$  & 5.7$^{+0.33}_{-0.43}$  & 3.32$^{+0.43}_{-0.43}$ & 0.66$^{+0.19}_{-0.09}$ \\
		2M17454705-2639109   &  -0.056      &  1   &      ...        & ... & ...  & ... & ... \\
	    2M17492967-2328298   &  0.117        &  2   &      0.10    & ... & ...  & ... & ... \\	
        2M17534571-2949362   &  ...              &   2  &      0.07    & 0.92$^{+0.89}_{-0.66}$& 6.18$^{+2.08}_{-0.97}$& 0.74$^{+1.74}_{-0.43}$ & 0.73$^{+0.18}_{-0.12}$\\		
		2M11462612-1419069   &  0.048        &  4   &      0.08    & ... & ...  & ... & ... \\
       2M17180311-2750124   &  0.039        &  2   &      0.08    & 0.167$^{+0.56}_{-0.12}$ & 5.40$^{+1.0}_{-1.59}$ & 2.27$^{+0.79}_{-0.68}$ & 0.94$^{+0.04}_{-0.19}$ \\ 		
       2M02491285+5534213   &  ...             &  3   &      0.20     & ... & ...  & ...  & ... \\
		\hline
		\hline
	\end{tabular} 
	\tablecomments{The orbital eccentricity is defined as $e= ( r_{apo} - r_{peri})/(r_{apo} + r_{peri})$, with $r_{apo}$ and $r_{peri}$ the perigalactic and apogalactic radii of the orbit, respectively. The orbital elements given here are estimates from Monte Carlo simulations of 10$^5$ orbits. }
\end{table*}

\section*{Acknowledgements}

J.G.F-T would like to dedicate this work in memory of his father, Jos\'e Gregorio Fern\'andez Rangel (1959-2017).

We are grateful to the referee for a prompt and constructive report. J.G.F-T is especially grateful for the technical expertise and assistance provided by the Instituto de Astrof\'isica de Canarias (IAC). J.G.F-T, D.G, and B.T gratefully acknowledge support from the Chilean BASAL Centro de Excelencia en Astrof\'isica y Tecnolog\'ias Afines (CATA) grant PFB-06/2007. S.V gratefully acknowledges the support provided by Fondecyt reg. n. 1170518. D.A.G.H. was funded by the Ram\'on y Cajal fellowship number RYC-2013-14182. D.A.G.H. and O.Z. acknowledge support provided by the Spanish Ministry of Economy and Competitiveness (MINECO) under grant AYA-2014-58082-P. Szabolcs M{\'e}sz{\'a}ros has been supported by the Premium Postdoctoral Research Program of the Hungarian Academy of Sciences, and by the Hungarian NKFI Grants K-119517 of the Hungarian National Research, Development and Innovation Office. D.M. is supported also by FONDECYT No. 1170121 and by the Ministry of Economy, Development, and Tourism's Millennium Science Initiative through grant IC120009, awarded to the Millennium Institute of Astrophysics (MAS). E.M, B.P, and A.P.V acknowledge support from UNAM/PAPIIT grant IN105916 and  IN 114114.

Funding for the \textit{GravPot16} software has been provided by the Centre national d'\'etudes spatiales (CNES) through grant 0101973 and UTINAM Institute of the Universit\'e de Franche-Comt\'e, supported by the Region de Franche-Comt\'e and Institut des Sciences de l'Univers (INSU).



Funding for the Sloan Digital Sky Survey IV has been provided by the Alfred P. Sloan Foundation, the U.S. Department of Energy Office of Science, and the Participating Institutions. SDSS- IV acknowledges support and resources from the Center for High-Performance Computing at the University of Utah. The SDSS web site is www.sdss.org. SDSS-IV is managed by the Astrophysical Research Consortium for the Participating Institutions of the SDSS Collaboration including the Brazilian Participation Group, the Carnegie Institution for Science, Carnegie Mellon University, the Chilean Participation Group, the French Participation Group, Harvard-Smithsonian Center for Astrophysics, Instituto de Astrof\`{i}sica de Canarias, The Johns Hopkins University, Kavli Institute for the Physics and Mathematics of the Universe (IPMU) / University of Tokyo, Lawrence Berkeley National Laboratory, Leibniz Institut f\"{u}r Astrophysik Potsdam (AIP), Max-Planck-Institut f\"{u}r Astronomie (MPIA Heidelberg), Max-Planck-Institut f\"{u}r Astrophysik (MPA Garching), Max-Planck-Institut f\"{u}r Extraterrestrische Physik (MPE), National Astronomical Observatory of China, New Mexico State University, New York University, University of Notre Dame, Observat\'{o}rio Nacional / MCTI, The Ohio State University, Pennsylvania State University, Shanghai Astronomical Observatory, United Kingdom Participation Group, Universidad Nacional Aut\'{o}noma de M\'{e}xico, University of Arizona, University of Colorado Boulder, University of Oxford, University of Portsmouth, University of Utah, University of Virginia, University of Washington, University of Wisconsin, Vanderbilt University, and Yale University.


\clearpage



\noindent \hrulefill

\noindent 
$^{1}$Departamento de Astronom\'ia, Universidad de Concepci\'on, Casilla 160-C, Concepci\'on, Chile\\
${^2}$Instituto de Astrof\'{\i}sica de Canarias, 38205 La Laguna, Tenerife, Spain\\
${^3}$Departamento de Astrof\'{\i}sica, Universidad de La Laguna, 38206 La Laguna, Tenerife, Spain\\
$^{4}$Institut Utinam, CNRS UMR 6213, Universit\'e Bourgogne-Franche-Comt\'e, OSU THETA Franche-Comt\'e, Observatoire de Besan\c{c}on, BP 1615, 25010 Besan\c{c}on Cedex, France\\
$^{5}$Observat\'orio Nacional, Rua Gal. Jos\'e Cristino 77, 20921-400 Rio de Janeiro, Brazil\\
${^6}$Astrophysics Research Institute, Liverpool John Moores University, 146 Brownlow Hill, Liverpool, L3 5RF, United Kingdom\\
$^{7}$New Mexico State University, Las Cruces, NM 88003, USA\\
$^{8}$Steward Observatory, University of Arizona, 933 North Cherry Avenue, Tucson, AZ 85721, USA\\
$^{9}$University of Texas at Austin, McDonald Observatory, Fort Davis, TX 79734, USA\\
$^{10}$Centro de Investigaciones de Astronom\'ia, AP 264, M\'erida 5101-A, Venezuela\\
$^{11}$Space Telescope Science Institute, 3700 San Martin Dr, Baltimore, MD 21218, USA\\
$^{12}$Department of Physics \& Astronomy, University of Utah, 115 S 1400 E, Salt Lake City, UT 84112, USA\\
$^{13}$Department of Astronomy, University of Virginia, Charlottesville, VA 22903, USA\\
$^{14}$Department of Physics and JINA Center for the Evolution of the Elements, University of Notre Dame, Notre Dame, IN 46556, USA\\
$^{15}$ELTE Gothard Astrophysical Observatory, H-9704 Szombathely, Szent Imre Herceg st. 112, Hungary\\
$^{16}$Premium Postdoctoral Fellow of the Hungarian Academy of Sciences\\
$^{17}$Leibniz-Institut f\"ur Astrophysik Potsdam (AIP), An der Sternwarte 16, 14482 Potsdam, Germany\\
$^{18}$Laborat\'orio Interinstitucional de e-Astronomia, - LIneA, Rua Gal. Jos\'e Cristino 77, Rio de Janeiro, RJ - 20921-400, Brazil\\
$^{19}$Departamento de Fisica, Facultad de Ciencias Exactas, Universidad Andres Bello, Av. Fernandez Concha 700, Las Condes, Santiago, Chile\\
$^{20}$Universidade Federal do Rio Grande do Sul, Instituto de Fisica, Av. Bento Goncalves 9500, Porto Alegre, RS, Brazil\\
$^{21}$Instituto Milenio de Astrofisica, Santiago, Chile\\
$^{22}$Vatican Observatory, V00120 Vatican City State, Italy\\
$^{23}$Instituto de Astrof\'isica, Pontificia Universidad Cat\'olica de Chile, Av. Vicuna Mackenna 4860, 782-0436 Macul, Santiago, Chile\\
$^{24}$Instituto de Astronom\'ia, Universidad Nacional Aut\'onoma de M\'exico, Apdo. Postal 70264, M\'exico D.F., 04510, M\'exico\\
$^{25}$Max-Planck-Instit\"ut f\"ur Extraterrestrische Physik, Gie\ss enbachstra\ss e, 85748 Garching, Germany\\
$^{26}$Laboratoire Lagrange, Universit\'e C\^ote d'Azur, Observatoire de la C\^ote d'Azur, CNRS, Blvd de l'Observatoire, F-06304 Nice, France\\
$^{27}$Departamento de F\'isica, Facultad de Ciencias, Universidad de La Serena, Cisternas 1200, La Serena, Chile\\
$^{28}$Unidad de Astronom\'ia, Universidad de Antofagasta, Avenida Angamos 601, Antofagasta 1270300, Chile\\
$^{29}$Department of Astronomy, Case Western Reserve University, Cleveland, OH 44106, USA\\
$^{30}$Apache Point Observatory and New Mexico State University, P.O. Box 59, Sunspot, NM, 88349-0059, USA\\
$^{31}$Sternberg Astronomical Institute, Moscow State University, Moscow, Russia\\
$^{32}$University of Utah Department of Physics and Astronomy 115 S. 1400 E., Salt Lake City, UT 84112, USA\\
$^{33}$Harvard-Smithsonian Center for Astrophysics, 60 Garden Street, Cambridge, MA 02138, USA\\
$^{34}$Instituto de Astronom\'ia, Universidad Nacional Aut\'onoma de M\'exico, Unidad Acad\'emica en Ensenada, Ensenada 22860, Mexico
\end{document}